\documentclass[12pt,preprint]{aastex61}
\usepackage{color}

\shortauthors{Okamoto and Sakurai}
\shorttitle{Super-strong sunspot magnetic field}
\begin{document}

\title{Super-strong Magnetic Field in Sunspots}

\correspondingauthor{Takenori J. Okamoto}
\email{joten.okamoto@nao.ac.jp}

\author[0000-0003-3765-1774]{Takenori J. Okamoto}
\altaffiliation{NAOJ fellow}
\affil{National Astronomical Observatory of Japan, Mitaka, Tokyo 181-8588, Japan}

\author[0000-0002-6019-5167]{Takashi Sakurai}
\affiliation{National Astronomical Observatory of Japan, Mitaka, Tokyo 181-8588, Japan}

\date{\today}

%------------------------------------------------------------------------------------------
%--- Abstract (ApJ max 250 words) -------------------------------------------------------------------------------
%------------------------------------------------------------------------------------------
\begin{abstract}

Sunspots are the most notable structure on the solar surface with strong magnetic fields. The field is generally strongest in a dark area (umbra), but sometimes stronger fields are found in non-dark regions such as a penumbra and a light bridge. The formation mechanism of such strong fields outside umbrae is still puzzling. Here we report clear evidence of the magnetic field of 6,250~G, which is the strongest field among Stokes~{\it I} profiles with clear Zeeman splitting ever observed on the Sun. The field was almost parallel to the solar surface and located in a bright region sandwiched by two opposite-polarity umbrae. Using a time series of spectral datasets, we discussed the formation process of the super-strong field and suggested that this strong field region was generated as a result of compression of one umbra pushed by the horizontal flow from the other umbra, like the subduction of the Earth's crust in plate tectonics.

\end{abstract}

\keywords{sunspots, Sun: magnetic fields, Sun: photosphere}

%------------------------------------------------------------------------------------------
%--- Introduction -----------------------------------------------------------------------
%------------------------------------------------------------------------------------------
\section{Introduction}

Sunspots are concentrations of magnetic fields on the solar surface. Their strong magnetic field controls the physical conditions in and around sunspots and produces various kinds of structures. For example, a sunspot usually consists of the umbra with vertical magnetic field and the penumbra with horizontal field \citep{bor11,rem09}. The penumbra harbors an outward flow of gas along the horizontal threads with a speed of several km~s$^{-1}$, which is called the Evershed flow \citep{eve09}. In mature umbrae, we can see bright cracks (light bridges), which have weaker fields than the surrounding umbrae. The light bridges are considered as convective cells penetrating from below the umbrae \citep{vaz73, jur06} and finally break up the sunspots \citep{bra64}. This assertion is also supported by the values of filling factor representing the ratio of a magnetized component in each observed pixel. Umbrae and penumbrae generally have large filling factors (i.e., close to unity), which means that the magnetized component almost entirely covers the pixels. On the other hand, granules outside sunspots show small filling factors \citep{oro07}, because of the dominance of non-magnetized gas. Light bridges also show smaller values than the surrounding umbrae \citep{lek97}, thus light bridges include a significant fraction of non-magnetized gas.

The darkness of umbrae is generally correlated with their magnetic field strength \citep{kin34, sch14}. Hence, the strongest magnetic field in each sunspot is located in the umbra in most cases \citep{sol03}. A typical field strength in sunspots is around 3,000~G \citep{rez12,liv15}, while \citet{liv06} reported 6,100~G among statistical data taken from 1917 through 2004. However, some exceptions also have been found outside umbrae. \citet{tan91} and \citet{zir93} found a strength of 4,300~G in complex sunspots with light bridges that separated opposite polarity umbrae. Interestingly, such kind of strong fields is nearly parallel to the solar surface, which is as strong as or much stronger than vertical umbral fields \citep{jae16}. As the strongest magnetic field ever reported, \citet{van13} showed $\sim$7,500~G in a sunspot penumbra with complex inversion technique. Although the proper motion of sunspots or flows in light bridges and penumbrae might contribute to the enhancement of horizontal fields, there is no convincing explanation about the formation mechanism of these strong fields. The origin and behavior of strong fields are also important for understanding various solar activities such as flares, mass ejections, flux ropes, and coronal heating.

Here we report an extremely strong magnetic field in a sunspot. Using the Solar Optical Telescope (SOT) aboard \emph{Hinode} \citep{kos07, tsu08, sue08, ich08, shi08}, we performed continuous observations of an active region to take full Stokes profiles (polarization profiles) by the Spectro-Polarimeter\footnote{Level 1 \emph{Hinode}/SP data, doi:10.5065/D6T151QF.} \citep[SP;][]{lit13} of the SOT. We investigated the time evolution and the spatial structure of the sunspot. We present the properties and discuss the formation mechanism of the strong field in this Letter.

%------------------------------------------------------------------------------------------
%--- Observation, reduction, selection---------------------------------------------
%------------------------------------------------------------------------------------------
\section{Observations and data reduction}

The \emph{Hinode} satellite observed an active region NOAA 11967 from 2014 February 1 to 6. We had 31 raster scans with the SP to obtain maps of the active region. The scanning was mainly performed with the Fast Mapping mode, which has an integration time of 3.2~s at each slit position and a pixel sampling of 0\farcs32. The field of view was 280\arcsec\ by 130\arcsec\ (200~Mm by 90~Mm on the Sun). The SP simultaneously measured the full Stokes profiles of the Fe~{\sc i}~lines at 6301.5~\AA\ and 6302.5~\AA\ with a sampling of 21.6~m\AA\ and with polarization sensitivity of 10$^{-3}$ relative to the continuum intensity.

Vector magnetic field, Doppler velocity, and filling factor were derived from the calibrated Stokes profiles under the assumption of a Milne-Eddington (ME) atmosphere. The inversion\footnote{Level~2 \emph{Hinode}/SP data, doi:10.5065/D6JH3J8D.} was performed with the MERLIN code \citep{bru07} developed under the Community Spectropolarimetric Analysis Center at the High Altitude Observatory (CSAC/HAO). The inversion operations are limited to a maximum field strength of 5,000~G and we found numerous pixels with 5,000~G in most of the raster scans. Hence, we also applied the MEKSY code\footnote{This code is available at: \texttt{http://hinode.nao.ac.jp/SDAS/manual/meksy\_Man\_E/index.html}.} developed at the National Astronomical Observatory of Japan to such pixels to derive the actual field strength beyond 5,000~G. The 180$^{\circ}$ azimuth ambiguity was resolved by the AZAM utility \citep{lit95}, where the basic premise is minimization of spatial discontinuities in the field orientation.

%We note that the field strengths are rounded at 10~G which is also roughly the accuracy of the measurements in this Letter.

%------------------------------------------------------------------------------------------
%--- Analysis ------------------------------------------------------------------------
%------------------------------------------------------------------------------------------

\section{Analysis}

\begin{figure}
\epsscale{1.2}
\plotone{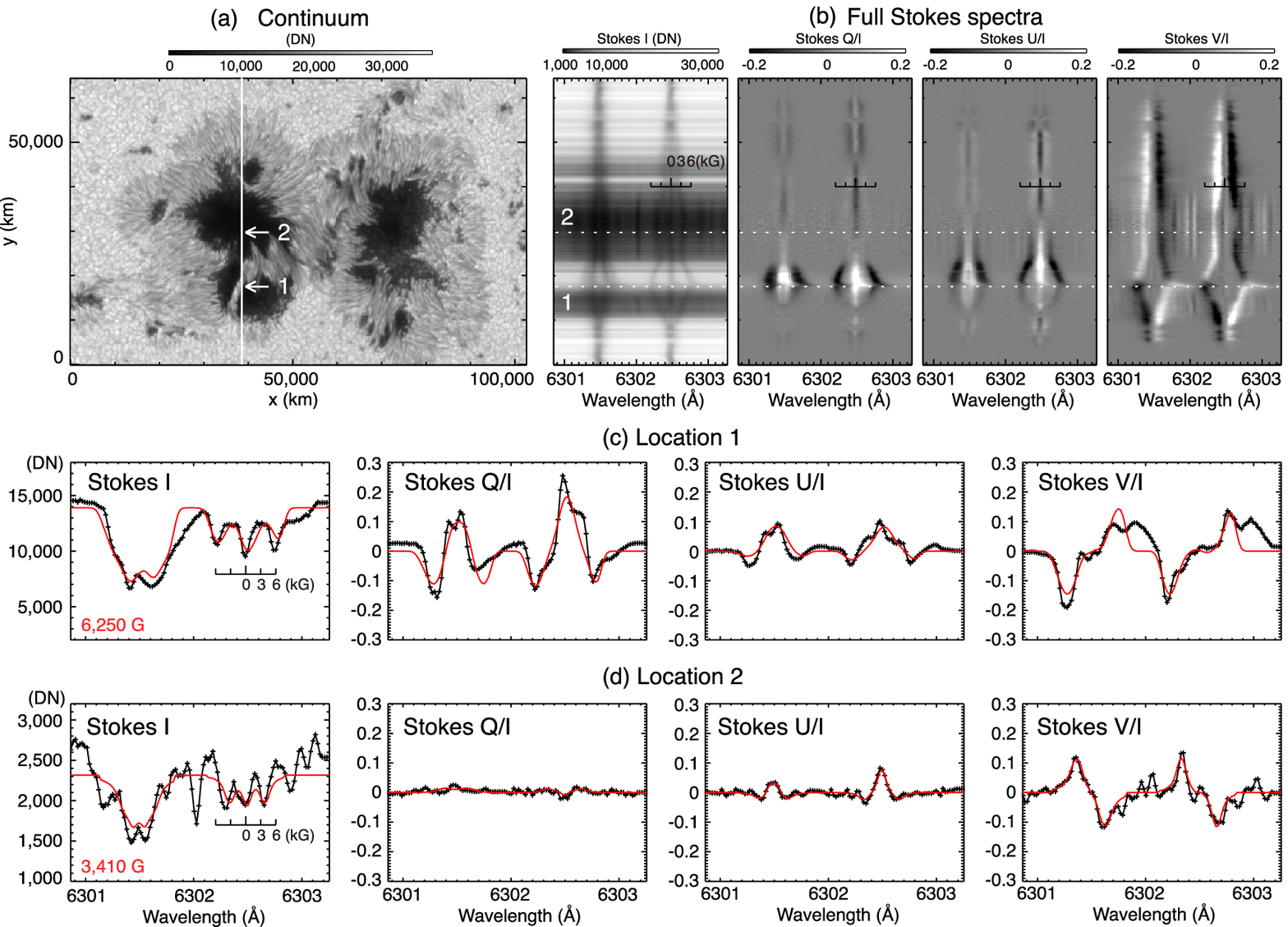}
\caption{
Sunspot and its spectra observed with the SOT/SP. (a) A continuum map of the sunspot scanned around 19~UT on 2014 February 4. North is up and east is to the left. (b) The full Stokes spectra at the slit position shown with the white line in (a). (c-d) Examples of observed Stokes profiles (black lines and crosses) and best-fit ones (red lines). Locations~1 and 2 represent the light bridge and the umbra indicated in (a-b). The numbers in red show the magnetic field strength derived by the MEKSY inversion. The scale for the Zeeman splitting (in kG) is shown in (b-d).
}
\label{fig1}
\end{figure}

The sunspot had a light bridge that divided the umbra into the northern and southern parts. An example of spectra along the white line (the slit position of SOT) in Figure~\ref{fig1}(a) is shown in Figure~\ref{fig1}(b). The magnetic field strength in the northern umbra was 3,500--4,500~G derived from the Zeeman splitting of the spectra, while the splitting drastically widens in the light bridge to exceed 6,000~G. We selected two locations indicated by arrows in Figure~\ref{fig1}(a) and compared the full Stokes profiles (black lines and crosses in Figure~\ref{fig1}(c-d)). The Zeeman splitting at location~1 is clear enough to measure the field strength easily without the need of any inversion techniques \citep{lan04}. However, some profiles outside the light bridge include molecular lines formed in lower temperature regions, in particular around the center of the umbra (e.g., location~2 in Figure~\ref{fig1}). Hence, we used the ME inversion of Stokes profiles for derivation of field strength in the entire field of view to reduce human biases and obtain other physical information as described in the previous section. The best-fit profiles by MEKSY are shown by red lines in Figure~\ref{fig1}. The maximum field strength in our observations was 6,250~G at Location~1, which consisted of 6,190~G and 860~G as horizontal and vertical components on the local frame, respectively. We note that the Stokes~{\it I}, {\it Q}, {\it U}, and {\it V} profiles at Location~1 indicate two magnetic components of the same polarity, one of which is strongly redshifted. The best-fit profiles by the ME inversion do not produce the extended component, but we can assure that the fitted profiles present the existence of the super-strong fields in the sunspot.

We investigated the time evolution and the spatial structure of the strong field region. Magnetic fields exceeding 4,000~G were located almost only in the northern part of the light bridge at the beginning of our observations (frames~1--4 of Figure~\ref{fig2}). Moreover, those exceeding 5,000~G (yellow contours) existed only at the boundary between the northern umbra and the light bridge. The light bridge apparently had no clear structures, but it had an elongated thread-like pattern running from northwest to southeast.

The strong field region gradually decayed (frames~5--6) and the elongated structure also changed and ended up like the penumbral threads running in the north-south direction. However, a new region with extremely strong fields exceeding 5,000~G appeared to the south of the preexisting strong region (frames~7--8). It was also on the light bridge but was widely spread in area. The region with field strength exceeding 6,000~G (red contour) was located at the southern boundary of the light bridge (frame~8). Then the field strength in the light bridge returned (frames~9--10) nearly to the level of the initial state.

We took a closer look at the strong field region to study the relationship between vector magnetic fields and Doppler velocity in and around the light bridge, and found five crucial features as follows. First, the two umbrae divided by the light bridge (Figure~\ref{fig3}(a)) had opposite polarities (orange and green in Figure~\ref{fig3}(b)). The polarity inversion line was located at the central axis of the light bridge, as expected for a delta spot. Second, the light bridge was filled with strong horizontal magnetic fields (Figure~\ref{fig3}(b), black bars). Third, the Doppler velocities showed blueshift along the horizontal fields in the light bridge (Figure~\ref{fig3}(c)). Fourth, strong redshifted motions were detected only at the locations where the horizontal fields and the umbral boundary crossed perpendicularly (Figure~\ref{fig3}(c--f)). Last, in the regions of strong redshift, the inclination of the magnetic field was much larger than the field inclination in the blueshift regions, so was the magnetic field strength (Figure~\ref{fig3}(g)).

Figure~\ref{fig4} shows the magnetic filling factor distributions in the sunspot region. The light bridge that we focused here shows large filling factors. The values are unity over a large area of the region as shown in red. On the other hand, we can see another bright structure in the southern umbra, which shows smaller filling factors than the surrounding umbra. Hence, the southern bright structure is considered as a typical light bridge, while the region we focused in this Letter is not a light bridge.

%------------------------------------------------------------------------------------------
%--- Discussion ------------------------------------------------------------------------
%------------------------------------------------------------------------------------------

\section{Discussion}

\subsection{The strongest field}

We observed super-strong fields in sunspots with clear Zeeman splitting in Stokes~{\it I} profiles. The strongest field in our observations was 6,250 G, which was located in the bright region sandwiched between two opposite-polarity umbrae. This is one of the strongest magnetic fields ever observed after the discovery of magnetic field on the Sun in 1908 \citep{hal08}. In particular, we can conclude that the horizontal component (6,190~G) of the field is the largest transverse magnetic field observed on the Sun. \citet{liv06} reported an umbral field with more than 6,000~G by measuring the Zeeman splitting only in Stokes~{\it I} spectra. We are aware that a recent observation with complex inversion techniques inferred a strong magnetic field ($\sim$7,500~G) in a sunspot \citep{van13}. The observed Stokes~{\it I} spectra, however, did not show clear Zeeman splitting, but broad absorption profiles. They interpreted that the very high values of the magnetic field strength are predominantly based on the very broad wings of the Stokes~{\it V} profiles, which can only be produced by a strong magnetic field near optical depth unity. Our case also shows a redshift excess in Stokes~{\it V/I} that was not fitted by the Milne-Eddington inversion as well as an absorption feature in the red wing in Stokes~{\it I} (Figure~\ref{fig1}). This indicates potential existence of a component with much larger strength. However, the profiles may consist of multiple components of differing Doppler velocities along the line of sight. We note that we are careful about the derivation of such a component in further analyses.

\subsection{Formation mechanism of the super-strong field}

Here we discuss the mechanism to form the super-strong field in the sunspot. A straightforward interpretation is an emerging flux, since sunspots and active regions are always formed by emerging magnetic flux coming from the solar interior \citep{zwa85}. In the early phase of emergence, horizontal fields appear on the photosphere first as a blueshifted structure with a rising speed of about 1~km~s$^{-1}$ \citep{lit98, che10}. Then both ends of the horizontal fields migrate away from the emergence zone to form two magnetic concentrations with opposite polarities such as plages and umbrae. During the emergence, the mass inside the flux flows down along the inclined flux tube and goes back to the photosphere at both footpoints of the flux \citep{bru69, kaw76}. At this moment, a redshift is observed. This is a common phenomenon on the Sun. That is, if a new flux emerges in between the two opposite-polarity umbrae, the apparent features would be similar to those described in the previous section. In this scenario, if the overlying magnetic fields of the active region impeded the ascent of the emerging flux without causing a flaring activity \citep{kus12}, gradual compression of the field lines in the photosphere could explain the strong fields.

%In addition, the pressure difference between the two ends with different field strengths would produce a siphon flow from the weaker-field to the stronger-field footpoints, which can be seen as a coherent flow along the horizontal fields.

However, we show two inconsistencies of the appearance in a typical flux emergence and our observations. The first one is about the duration of the blueshift in the penumbra. We see the coherent blueshifted structure along the penumbra with a line-of-sight velocity of 1--3~km~s$^{-1}$. It always existed for five days from the start to the end of our observations. Therefore, we cannot support a theory that the continuous blueshift is caused by a rising motion of a single magnetic flux, because of the too-long duration for the compact area ($\sim$30,000~km). Multiple emergence may be another solution, but we can exclude this possibility as well, since the configuration of the penumbra did not change drastically in a short time. The second inconsistency is about the center-to-limb variation of the Doppler velocity. We point out that the velocity was larger when the sunspot was far from the disk center, and smaller when close to the disk center on February 3 (Figure~\ref{fig3}). In the case of flux emergence, the apparent rising velocity must have been larger at the location close to the disk center, because of the projection effect. 
%On the other hand, if we assume that the blueshift is a projected velocity of a horizontal flow along the penumbra, the actual speed is estimated to be an almost constant value (6.5--7.2~km~s$^{-1}$) for five days. %Hence, this assumption is more convincing and we conclude that the strongest field was not a part of an emerging flux, but it was generated by the surface motion as shown in Figure~\ref{fig5}.

%{\bf 
%Considering the Doppershift patterns along the bright structure, one may presume that the velocity features indicate a siphon flow. The redshifts were always located in regions of stronger fields than the blueshifts (Figure~\ref{fig3}), which is consistent with the travel direction of a siphon flow driven by the pressure difference between the two ends with different field strengths. One possible way to prepare such a magnetic configuration is magnetic reconnection between the neighboring umbrae. The reconnected fields would retract due to magnetic tension and be compressed up to the observed field strength. We note that we do not exclude this scenario, but unfortunately it is difficult to verify it only with our observations. The generation mechanism of the asymmetric Dopplershifts along the bright feature and the relationship with flares must be clarified.
%}

Hence, we suggest an alternative scenario to interpret the observed phenomena as follows.\footnote{The referee has suggested another explanation as follows. The blueshifts might be caused by a siphon flow mechanism. The strong field was generated due to magnetic reconnection between the two umbrae . The phenomena associated with the reconnection combined with flux emergence may work towards increasing the field strength to the observed values, by compressing the fields.} The bright region filled with strong horizontal fields was actually part of the penumbra of the southern umbra rather than belonging to the light bridge. This assertion is also supported by a high filling factor (almost unity) in the bright region, derived from the ME inversion; the light bridges usually show low filling factors \citep{lek97}. The blueshifted motions occupying the bright region can be considered as the line-of-sight component of the horizontal (north-westward) flow along the penumbral threads, since the sunspot was located in the southern hemisphere (about 7$^{\circ}$ to the south from the disk center). Under this assumption, the speed of the field-aligned horizontal flow was estimated to be 7.2~km~s$^{-1}$, which is consistent with a typical speed of the Evershed flow \citep{bel07}. Both the northern and southern umbrae were supposed to have attempted to form their own penumbrae in the buffer area, but the southern one dominated at the beginning of our observations. The flow prevented the northern umbra from forming its penumbra on its southern side. The strength of the horizontal field was not large compared to that of the umbral fields, as long as the orientations of the field and the horizontal\footnote{The referee has claimed that the flows are not Evershed flows, although we prefer to call them Evershed flows. Thus we avoid the wording here.} flow were parallel to the umbral boundaries (Figure~\ref{fig3}(b)). However, the front of the penumbra from the southern umbra eventually reached the northern umbra and the northward horizontal flow pushed up on the umbral fields. At this moment, the umbral fields and nearby horizontal fields were compressed by the flow and enhanced at the boundary of the northern umbra, and the flow went downward there showing redshift, which reminds us of subduction of the crust in the Earth's plate tectonics (Figure~\ref{fig5}).

\begin{figure}
\epsscale{1.2}
\plotone{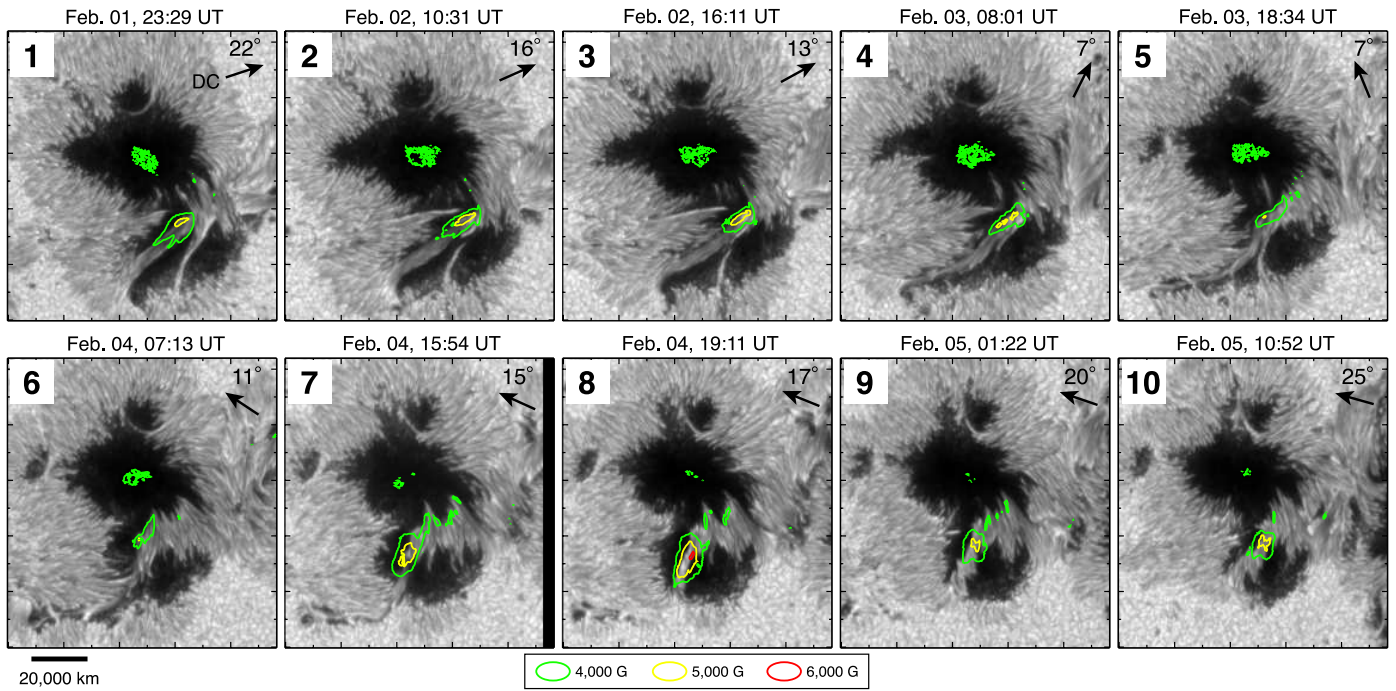}
\caption{
Time series of the continuum images of the sunspot. Contours indicate the magnetic field strength (green, yellow, and red for 4, 5, 6~kG). Each panel shows the direction of and the angular distance to the disk center at the center of the field of view.
}
\label{fig2}
\end{figure}

\begin{figure}
\epsscale{1.2}
\plotone{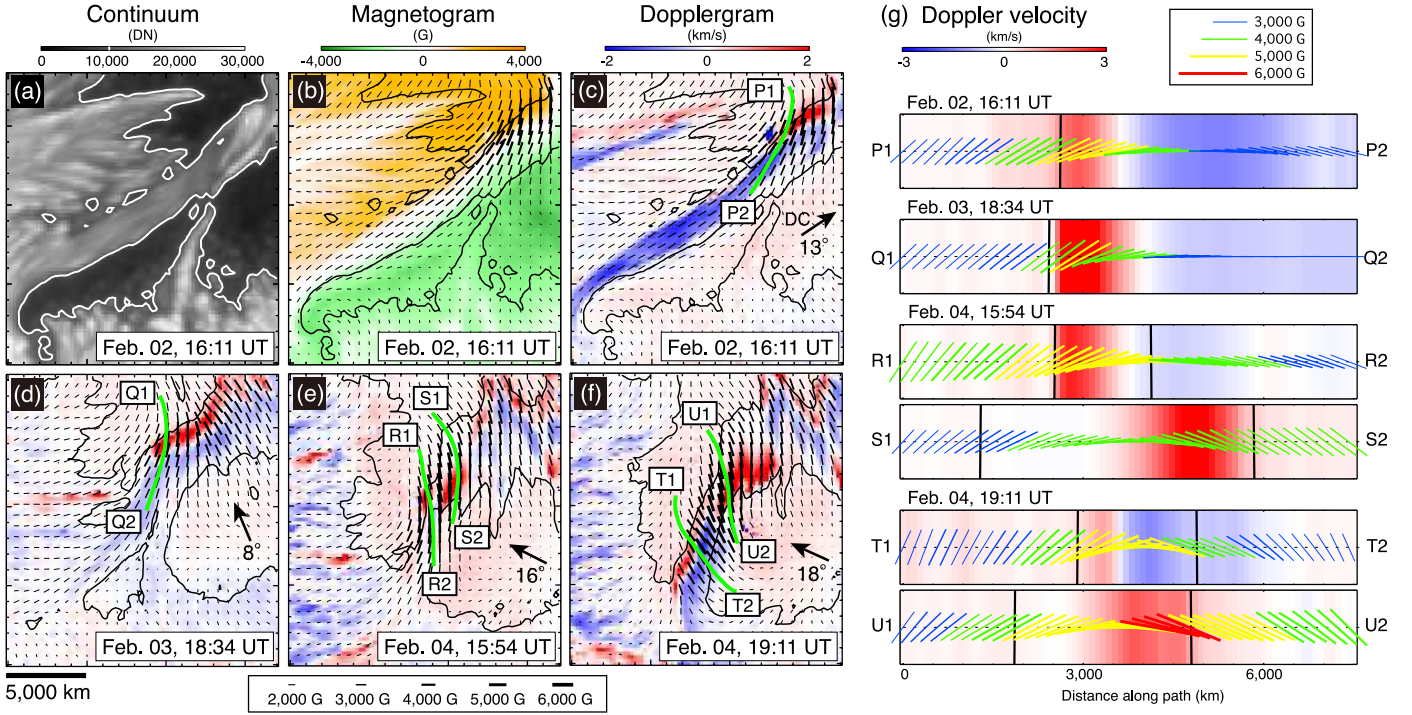}
\caption{
Vector magnetic fields and Doppler velocities in the sunspot region. (a) A continuum map. The white contours indicate the boundary of the umbrae. The same contours are also drawn in black in (b) and (c). (b) A vector magnetogram. The color background (orange for positive and green for negative polarities) and the black bars show the vertical and the horizontal components of the magnetic field in the solar local frame, respectively. (c--f) Time series of Doppler (line-of-sight) velocity maps (blue and red mean velocities toward and away from us). Velocities exceeding $\pm$2~km~s$^{-1}$ are saturated. The black bars are the horizontal magnetic fields as in (b). Panels (a--c) correspond to frame~3 of Figure~\ref{fig2}, while panels (d--f) correspond to frames~5, 7, and 8, respectively. Each panel shows the direction of and the angular distance to the disk center at the center of the field of view. (g) Horizontal profiles of vector magnetic fields and Doppler velocities along the green paths shown in (c--f). These green paths are drawn along the horizontal magnetic field vectors. The background color (from blue to red) indicates Doppler velocities. The color bars show the inclination and strength of magnetic field vectors. The vertical black lines represent the boundary of the umbrae.
}
\label{fig3}
\end{figure}

\begin{figure}
\epsscale{0.6}
\plotone{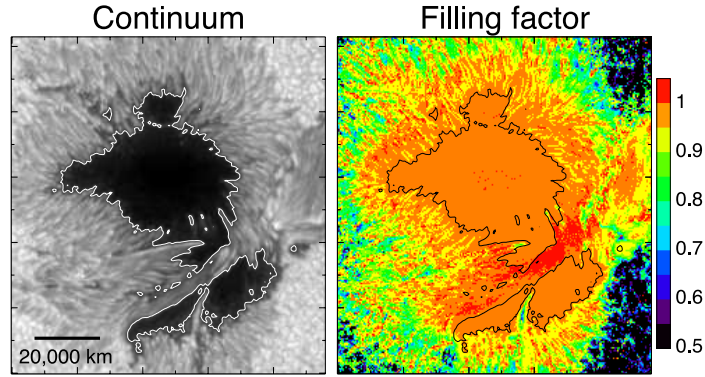}
\caption{
A continuum map and the corresponding filling factor map of the sunspot. The white/black lines indicate the umbral boundary.
}
\label{fig4}
\end{figure}

\begin{figure}
\epsscale{0.7}
\plotone{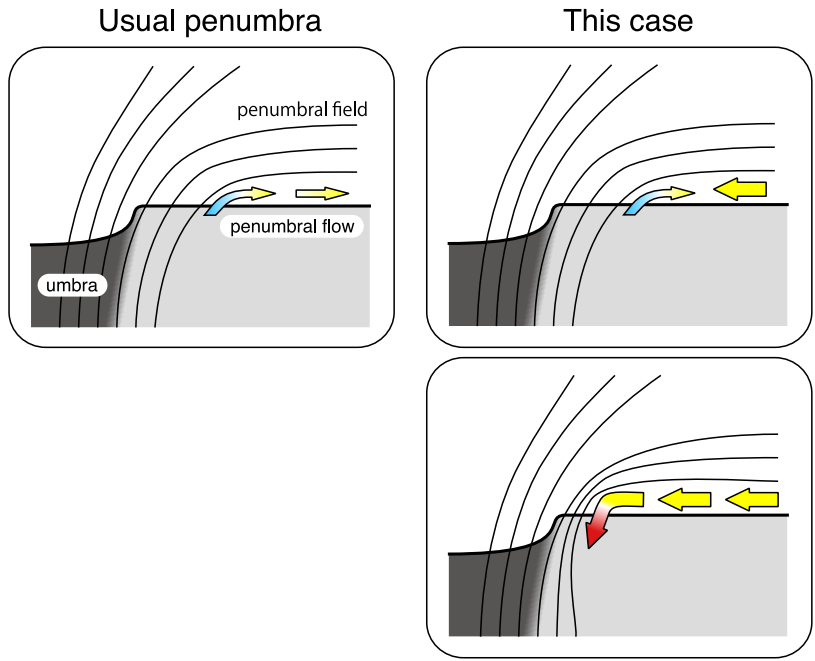}
\caption{
Schematic illustrations of the formation mechanism of strong magnetic fields near the umbral boundary. In a usual case, a penumbral flow goes outward from the sunspot umbra as shown in the left panel. If another flow generated by the other umbra comes from the opposite direction and is strong enough, they collide each other and the strong flow rolls back the weak one. Finally, the flow pushes up and enhance the magnetic field in the vicinity of the umbra.
}
\label{fig5}
\end{figure}

Combining these presumed steps with the observed time evolution of the spatial distribution of the strong fields (Figure~\ref{fig2}), we arrive at a comprehensive scenario on how the strongest field was formed. First, the umbral fields with 4,000~G were enhanced to 5,000~G by the process mentioned above (path P in Figure~\ref{fig3}(g)). Next, the enhanced field region gradually moved eastward, and the spatial configurations of the two umbrae also slightly changed. As a result, the penumbra from the southern umbra could not go toward northwest, but went toward northeast, which was the direction that the enhanced region existed. Hence, the enhanced umbral fields and related horizontal fields were further intensified to 5,500~G (path R in Figure~\ref{fig3}(g)). In addition, the northern umbra with the enhanced fields also got a chance to have its penumbra to extend toward south at this moment, and finally the enhanced field was compressed (like Figure~\ref{fig5} but now with the southern umbra) enough in the narrow region to have a strength of beyond 6,000~G (path U in Figure~\ref{fig3}(g)).

We examine the feasibility of the generation of the strong fields as a result of compression by the horizontal flow. Now let us compare the pressure balance between the flow and the magnetic pressure of the umbral fields. Using the mass density in sunspot umbrae \citep[1.1$\times$10$^{-6}$~g~cm$^{-3}$ in the model M of][]{mal86}, the magnetic field strength (5,000 G), and the flow speed (7~km~s$^{-1}$), we estimate that the magnetic pressure is comparable to the flow. A more precise analysis based on a realistic numerical simulation setup is needed to confirm or refute the results of the simple calculations.

%The process to enhance the magnetic fields in the sunspot requires a few repetitions of pushing by penumbral flows as well as configuration changes of the umbrae to promote collision of flows. Although the repetition of such processes was rare, similar dynamics and magnetic configurations were also seen in other sunspots with strong fields that we investigated (not shown here, but will be presented in our subsequent paper). 
Our proposed scenario provides a hint to understand complex motions and configuration changes in the penumbrae between two opposite-polarity umbrae reported so far. A single pixel of ground-based instruments such as the Advanced Stokes Polarimeter \citep[ASP;][]{elm92} may have included two or more components, which were interpreted as a mixture of horizontal flows and downward motions \citep{lit02}, but the 1\arcsec\ spatial resolution of the ASP may have been insufficient to distinguish them. Our high-resolution observations support this interpretation.

\acknowledgements 

We thank B. W. Lites, J. W. Harvey, W. Livingston, and an anonymous referee for their useful comments. We also thank T. Yokoyama, M. Kubo, Y. Katsukawa, M. Shimojo, and M. L. DeRosa for their technical supports. \emph{Hinode} is a Japanese mission developed and launched by ISAS/JAXA, with NAOJ as a domestic partner and NASA and STFC (UK) as international partners. It is operated by these agencies in cooperation with ESA and NSC (Norway). This work was supported by JSPS KAKENHI Grant Number 16K17663 (PI: T.J.O.) and 25220703 (PI: S. Tsuneta) and partly carried out on the Solar Data Analysis System operated by the Astronomy Data Center in cooperation with the Hinode Science Center of NAOJ.

%------------------------------------------------------------------------------------------
%--- Reference ------------------------------------------------------------------------------
%------------------------------------------------------------------------------------------

%------------------------------------------------------------------------------------------
%--- Figures --------------------------------------------------------------------------------
%------------------------------------------------------------------------------------------

\begin{thebibliography}{33}

\bibitem[Bellot Rubio et al.(2007)]{bel07} Bellot Rubio, L.~R., Tsuneta, S., Ichimoto, K., et al.\ 2007, \apjl, 668, L91 
\bibitem[Borrero \& Ichimoto(2011)]{bor11} Borrero, J.~M., \& Ichimoto, K.\ 2011, Living Reviews in Solar Physics, 8, 4 
\bibitem[Bray \& Loughhead(1964)]{bra64} Bray, R.~J., \& Loughhead, R.~E.\ 1964, Sunspots (London: Chapman \& Hall)  
\bibitem[Bruzek(1969)]{bru69} Bruzek, A.\ 1969, \solphys, 8, 29 
\bibitem[Cheung et al.(2010)]{che10} Cheung, M.~C.~M., Rempel, M., Title, A.~M., \& Sch{\"u}ssler, M.\ 2010, \apj, 720, 233 
\bibitem[Elmore et al.(1992)]{elm92} Elmore, D.~F., Lites, B.~W., Tomczyk, S., et al.\ 1992, \procspie, 1746, 22 
\bibitem[Evershed(1909)]{eve09} Evershed, J.\ 1909, \mnras, 69, 454 
%\bibitem[Frutiger et al.(2000)]{fru00} Frutiger, C., Solanki, S.~K., Fligge, M., \& Bruls, J.~H.~M.~J.\ 2000, \aap, 358, 1109 
\bibitem[Hale(1908)]{hal08} Hale, G.~E.\ 1908, \apj, 28, 315 
\bibitem[Ichimoto et al.(2008)]{ich08} Ichimoto, K., Lites, B., Elmore, D., et al.\ 2008, \solphys, 249, 233 
\bibitem[Jaeggli(2016)]{jae16} Jaeggli, S.~A.\ 2016, \apj, 818, 81 
\bibitem[Jur{\v c}{\'a}k et al.(2006)]{jur06} Jur{\v c}{\'a}k, J., Mart{\'{\i}}nez Pillet, V., \& Sobotka, M.\ 2006, \aap, 453, 1079 
\bibitem[Kawaguchi \& Kitai(1976)]{kaw76} Kawaguchi, I., \& Kitai, R.\ 1976, \solphys, 46, 125 
\bibitem[King(1934)]{kin34} King, R.~B.\ 1934, \apj, 80, 136 
\bibitem[Kosugi et al.(2007)]{kos07} Kosugi, T., Matsuzaki, K., Sakao, T., et al.\ 2007, \solphys, 243, 3 
\bibitem[Kusano et al.(2012)]{kus12} Kusano, K., Bamba, Y., Yamamoto, T.~T., et al.\ 2012, \apj, 760, 31 
\bibitem[Landi Degl'Innocenti \& Landolfi(2004)]{lan04} Landi Degl'Innocenti, E., \& Landolfi, M.\ 2004, Polarization in Spectral Lines (Dordrecht: Kluwer Academic Publishers, 307)
\bibitem[Leka(1997)]{lek97} Leka, K.~D.\ 1997, \apj, 484, 900 
\bibitem[Lites \& Ichimoto(2013)]{lit13} Lites, B.~W., \& Ichimoto, K.\ 2013, \solphys, 283, 601 
\bibitem[Lites et al.(1995)]{lit95} Lites, B.~W., Low, B.~C., Martinez Pillet, V., et al.\ 1995, \apj, 446, 877 
\bibitem[Lites et al.(1998)]{lit98} Lites, B.~W., Skumanich, A., \& Martinez Pillet, V.\ 1998, \aap, 333, 1053 
\bibitem[Lites et al.(2002)]{lit02} Lites, B.~W., Socas-Navarro, H., Skumanich, A., \& Shimizu, T.\ 2002, \apj, 575, 1131 
\bibitem[Lites et al.(2007)]{bru07} Lites, B., Casini, R., Garcia, J., \& Socas-Navarro, H.\ 2007, \memsai, 78, 148 
\bibitem[Livingston et al.(2006)]{liv06} Livingston, W., Harvey, J.~W., Malanushenko, O.~V., \& Webster, L.\ 2006, \solphys, 239, 41 
\bibitem[Livingston \& Watson(2015)]{liv15} Livingston, W., \& Watson, F.\ 2015, \grl, 42, 9185 
\bibitem[Maltby et al.(1986)]{mal86} Maltby, P., Avrett, E.~H., Carlsson, M., et al.\ 1986, \apj, 306, 284 
\bibitem[Orozco Su{\'a}rez et al.(2007)]{oro07} Orozco Su{\'a}rez, D., Bellot Rubio, L.~R., del Toro Iniesta, J.~C., et al.\ 2007, \apjl, 670, L61 
\bibitem[Rempel et al.(2009)]{rem09} Rempel, M., Sch{\"u}ssler, M., Cameron, R.~H., \& Kn{\"o}lker, M.\ 2009, Science, 325, 171 
\bibitem[Rezaei et al.(2012)]{rez12} Rezaei, R., Beck, C., \& Schmidt, W.\ 2012, \aap, 541, A60 
\bibitem[Schad(2014)]{sch14} Schad, T.~A.\ 2014, \solphys, 289, 1477 
%\bibitem[Scharmer et al.(2002)]{sch02} Scharmer, G.~B., Gudiksen, B.~V., Kiselman, D., L{\"o}fdahl, M.~G., \& Rouppe van der Voort, L.~H.~M.\ 2002, \nat, 420, 151 
\bibitem[Shimizu et al.(2008)]{shi08} Shimizu, T., Nagata, S., Tsuneta, S., et al.\ 2008, \solphys, 249, 221 
\bibitem[Solanki(2003)]{sol03} Solanki, S.~K.\ 2003, \aapr, 11, 153 
%\bibitem[Stix(2002)]{sti02} Stix, M.\ 2002, The sun : an introduction -- 2nd ed.~/Michael Stix.~Berlin : Springer, 2002.~QB 521 .S75,  
\bibitem[Suematsu et al.(2008)]{sue08} Suematsu, Y., Tsuneta, S., Ichimoto, K., et al.\ 2008, \solphys, 249, 197 
\bibitem[Tanaka(1991)]{tan91} Tanaka, K.\ 1991, \solphys, 136, 133 
\bibitem[Tsuneta et al.(2008)]{tsu08} Tsuneta, S., Ichimoto, K., Katsukawa, Y., et al.\ 2008, \solphys, 249, 167 
\bibitem[van Noort et al.(2013)]{van13} van Noort, M., Lagg, A., Tiwari, S.~K., \& Solanki, S.~K.\ 2013, \aap, 557, A24 
\bibitem[Vazquez(1973)]{vaz73} Vazquez, M.\ 1973, \solphys, 31, 377 
\bibitem[Zirin \& Wang(1993)]{zir93} Zirin, H., \& Wang, H.\ 1993, \solphys, 144, 37 
\bibitem[Zwaan(1985)]{zwa85} Zwaan, C.\ 1985, \solphys, 100, 397 

\end{thebibliography}
\end{document}